\documentclass{edp-jp4}
\usepackage{graphicx}
\begin{document}

\title{Recent results on energy relaxation in disordered charge
and spin density waves} 
\author{R. M\'elin}\address{Centre de Recherches sur les Tr\`es Basses
Temp\'eratures (CRTBT)\footnote{U.P.R. 5001 du CNRS, Laboratoire conventionn\'e
avec l'Universit\'e Joseph Fourier},\\ Bo\^{\i}te Postale 166,
F-38042 Grenoble Cedex 9, France}
\author{K. Biljakovi\'c}\address{Institute of Physics,
Hr-10 001 Zagreb, P.O. Box 304, Croatia}
\author{J. C. Lasjaunias}\sameaddress{1}
\author{P. Monceau}\sameaddress{1}

\maketitle
\begin{abstract} 
We briefly
review different approaches used recently to describe
collective effects
in the strong pinning model of disordered charge
and spin density waves, in connection with the CRTBT
very low temperature heat relaxation experiments.
\end{abstract}

\section{Introduction}

Extensive investigations of energy relaxation at very low temperature
(below $0.5$~K) in
disordered charge and spin density waves (CDWs and SDWs)
(see for instance \cite{Biljak89,Biljak91,Lasjau96,Lasjau02,Lasjau05})
have revealed ageing in the heat response, in the sense that the
temperature signal $T(t_w,\tau)$
measured at a time $t_w+\tau$ depends on
the ``waiting time'' (the duration $t_w$ of the heat perturbation),
as well as on time $\tau$ elapsed since $t_w$. Ageing
is ``interrupted'' \cite{Bouchaud}:
$T(t_w,\tau)$
does not depend on $t_w$ if $t_w$ is larger than
the maximal relaxation time
$\tau_{\rm max}$.

These properties, observed in a variety
of different compounds, are likely to originate from
the interplay between impurities that pin the phase of the CDW,
and the commensurate potential, as first discussed by Abe \cite{Abe}.
Fukuyama \cite{Fukuyama} discussed in a similar model
the possibility of a transition to a density wave glass.
More recently,
Larkin \cite{Larkin} and Ovchinikov \cite{Ovchinikov}
discussed metastability,
explaining why the CDW or SDW can absorb and restitute
energy over long time scales.
We provided recently a discussion of collective effects
in disordered CDWs and SDWs \cite{EPJB1,EPJB2}.

\section{Experimental results}
\label{sec:exp}
\begin{figure}
\begin{center}
\includegraphics [width=.8 \linewidth]{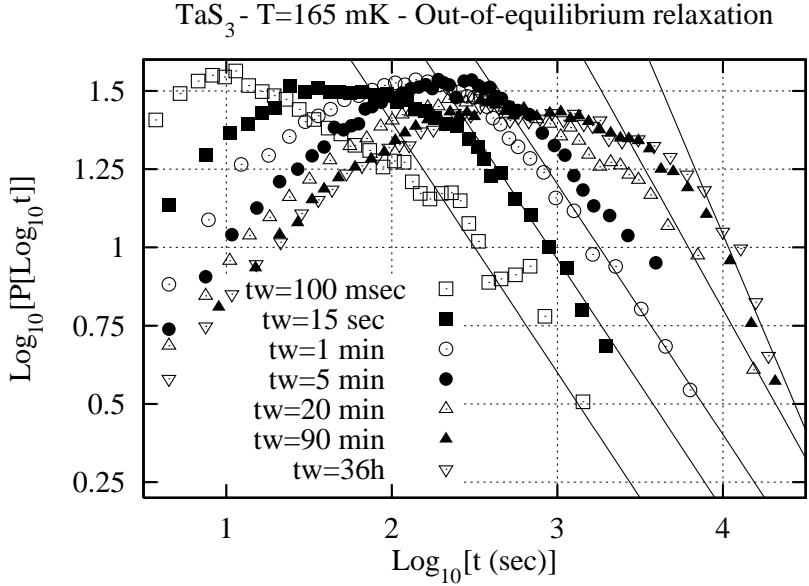}
\end{center}
\caption{Out-of-equilibrium
relaxation time spectrum of the
incommensurate CDW compound TaS$_3$ at the temperature
$T=165$~mK. Thermal equilibrium
has been reached for the waiting time $t_w=90$~min.
The long-time tail of the spectrum of
relaxation times is well fitted by
a power-law (solid lines). From Ref.~\cite{EPJB1}.
\label{fig:exp}
}
\end{figure}

Let us first summarize the experiments.
(i) As the temperature decreases, the total specific heat
first follows a $T^3$ behavior at
``high'' temperatures (due to phonons), followed by a minimum,
followed by a $T^{-2}$ behavior at low temperature.
(ii) The residual specific heat follows
with a power-law temperature dependence
once the $T^3$ and the $T^{-2}$ contributions have
been subtracted.
(iii) The amplitude of the $T^{-2}$ term decreases strongly
as the waiting time decreases.
(iv) The spectrum of relaxation times shows a 
power-law distribution for intermediate waiting times,
and interrupted ageing for larger waiting times (see Fig.~\ref{fig:exp}).
(v) Commensurate systems relax faster than incommensurate systems.

(i), (iii), (v) can be explained qualitatively by the properties
of independent strong pinning impurities \cite{Larkin,Ovchinikov}.
We explain
(iv) by collective effects in the strong pinning 
model \cite{EPJB1,EPJB2}.
We explain (ii) by collective effects for
substitutional disorder \cite{FM,EPJB2}.

\section{Phenomenological REM-like trap model}

The trap model inspired from the random energy model (REM) \cite{Derrida}
with an exponential distribution of trap energies
shows a divergence of the average relaxation time at the glass
temperature $T_g$ \cite{Bouchaud}.
The distribution $p(E)=\exp{(-E/T_g)}$ of the trap energies $E$,
combined to the
Aharenius time $\tau=\tau_0 \exp{(E/T)}$ (with
$\tau_0$ the ``microscopic'' time and $T$ the temperature),
leads to a power-law relaxation time spectrum
$p(\tau)\sim(\tau_0/\tau)^{1+T/T_g}$,
having an infinite first moment if the temperature
$T$ is lower than the glass transition temperature $T_g$.

We have shown \cite{EPJB1}
that heat relaxation experiments
in CDWs and SDWs can be well described phenomenologically
by a REM-like model 
with an exponential trap energy distribution and with an energy
cut-off $E_{\rm max}$. This model leads both to interrupted ageing with a
relaxation time $\tau_0 \exp{(E_{\rm max}/T)}$, and to a power-law
relaxation. The temperature $T_g$ in the REM-like model
relevant to CDWs and SDWs corresponds to
a cross-over temperature, with
no genuine divergence of the average
relaxation time.

\section{Independent strong pinning impurities}
The starting point is the local model of strong pinning
(see the recent review by Brazovskii and Nattermann \cite{revue}
and references therein),
defined by the Hamiltonian 
\begin{equation}
{\cal H}=\frac{\hbar v_F}{4\pi} \int dy 
\left( \frac{\partial \varphi(y)}{\partial y}\right)^2
+w\int dy \left[1-\cos{\varphi(y)}\right]
-\sum_i V_i \left[1-\cos{(Q y_i+\varphi(y_i))}\right]
,
\end{equation}
where $\varphi(y)$ is the phase of the CDW or SDW with
$y$ the coordinate along the chain, $v_F$ is the Fermi
velocity, $w$ the commensurate potential due to interchain
interactions and $V_i$ the pinning potential of the
impurity located at site $y_i$.
Metastability
due to bisolitons occurs for a sufficiently strong
impurity pinning potential \cite{Larkin,revue}, from what
it is possible to define an effective two-level system with
a ground state separated from a metastable state by an energy
barrier.
This
explains the $T^{-2}$ contribution to the specific heat,
as the high temperature
tail of a Schottky anomaly of the effective two level system
\cite{Larkin,Ovchinikov,EPJB2}.
A maximum in the specific heat is predicted at a temperature
even lower than the very low temperatures used in experiments.
Moreover, there exists for this model
a qualitative difference between commensurate
and incommensurate systems already for independent impurities
\cite{Lasjau05}:
in commensurate systems, the metastable state is degenerate with the
ground state, leading to a degenerate effective two-level system
that cannot absorb or restitute energy. This agrees qualitatively
with the experimental trend (v) in section~\ref{sec:exp}.

\section{Collective effects}
The power-law spectrum of relaxation times obtained in experiments
(see Fig.~\ref{fig:exp})
can hardly be explained by the dynamics of independent bisolitons.
In a first approximation, we consider
local deformations of the CDW induced by impurities, and
include the possibility that
several impurities can pin a given local deformation of the CDW.
Noting $\xi$ the width of the soliton, a given chain is divided in
a set of clusters, in such a way that two neighboring impurities belong
to the same cluster if they are at a distance smaller than $\xi$.
It can be shown \cite{EPJB2}
that this ``cluster'' model leads to the exponential energy
barrier distribution discussed previously
within the REM-like trap model. Moreover, 
the waiting time dependence of the energy relaxation can
be addressed within this model. As a weak point,
there exists a genuine glass transition and no interrupted
ageing, due to
the fact that the CDW is considered to be rigid in between two
impurities. 

\begin{figure}
\begin{center}
\includegraphics [width=.5 \linewidth]{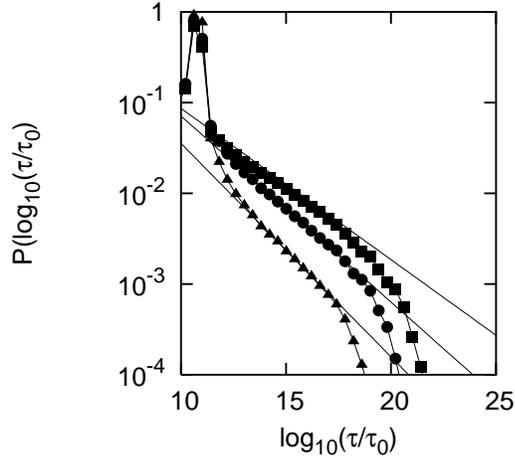}
\end{center}
\caption{Spectrum of relaxation times deduced from
dynamical renormalization group following a quench.
The different curves correspond to
different values of $x\xi$, where $x$ is the concentration of
impurities and $\xi$ the soliton width. We use
$x \xi=1.5$ (squares),
$x\xi=1$ (circles) and $x\xi=0.5$ (triangles).
From Ref.~\cite{EPJB2}.
\label{fig:rg}
}
\end{figure}

To include the deformations of the CDW in between two impurities
at an arbitrary distance, we simulated the evolution of a system of
randomly distributed bisolitons following a quench
by a dynamical renormalization group (RG) \cite{Fisher}, from what
we deduce the spectrum of relaxation times. We then obtain a
power-law relaxation and interrupted ageing, like in the
experiments and like in the REM-like trap model (see Fig.~\ref{fig:rg}).

\section{Substitutional impurities}
The model of substitutional disorder
proposed in Ref. \cite{FM} for a dimerized system
is quite different from the strong
pinning model. Substitutional disorder interpolates between edge states
obtained by cutting a chain, and solitons at the junction of two
distinct ground states.
We generalized Ref. \cite{FM} to substitutional
disorder in incommensurate systems \cite{EPJB2}.
The bound states generated by substitutional impurities
are exactly in the middle of the gap
for independent impurities. The degeneracy is lifted for two neighboring
impurities, with oscillations of the bound state levels in incommensurate
systems, and an overall exponential decay. Combining the exponential
distribution of the spacing between impurities to an exponential decay
of the effective hopping, leads to a power-law temperature dependence of
the specific heat, apparently observed in CDWs and SDWs [see (ii)
in section~\ref{sec:exp}].

\section{Conclusions}
To conclude, we have summarized recent results on energy relaxation in
disordered CDWs and SDWs, with an emphasis on collective effects,
both in the strong pinning model and for substitutional disorder.
This provides an explanation to several experimental observations.

\end{document}